\documentclass[aps,prd,twocolumn,superscriptaddress,floatfix,nofootinbib,longbibliography]{revtex4-2}
\usepackage{amsmath,amssymb,bm}
\usepackage{graphicx}
\usepackage[colorlinks=true,linkcolor=blue,citecolor=blue,urlcolor=blue]{hyperref}
\usepackage{orcidlink}
\usepackage[caption=false]{subfig}
\usepackage{bm}
\usepackage{xcolor}
\hypersetup{
	colorlinks=true,
	linkcolor=blue,
	citecolor=cyan,
	urlcolor=cyan
}
\begin{document}

\title{Schwinger instability for near-extremal charged BTZ black hole}

\author{Mendrit Latifi\,\orcidlink{0000-0003-0164-4396}}
\email{ml67500@student.uni-lj.si}
\affiliation{Faculty of Mathematics and Physics, University of Ljubljana, Jadranska ulica 19, SI-1000 Ljubljana, Slovenia}

\author{Kimet Jusufi\,\orcidlink{0000-0003-0527-4177}}
\email{kimet.jusufi@unite.edu.mk}
\affiliation{Physics Department, State University of Tetovo, Ilinden Street nn, 1200, Tetovo, North Macedonia}

\begin{abstract}
We study the near-horizon region of a near-extremal \emph{charged} BTZ black hole, where a controlled AdS$_2\times S^1$ throat emerges. In this background we solve the Klein–Gordon equation for a charged scalar exactly, identify the BF-bound threshold, and compute the mean Schwinger pair-production rate, which reduces to the standard semiclassical formula in the appropriate limit. The hypergeometric solution is exact within the decoupled near-horizon AdS$_2$ black-hole patch; with ingoing conditions imposed at the true outer horizon, it yields a closed-form mean number of produced pairs throughout the controlled near-extremal throat regime. Its finite-temperature dependence enters only through the ratio of the mode frequency to the Hawking temperature, while the extremal limit reproduces the known AdS$_2$ Schwinger result and admits an exact thermal-like form in terms of curvature-corrected effective temperatures. We also compute charged Hawking emission via a WKB tunneling analysis done directly in throat coordinates, allowing a clean comparison between thermal and electric contributions and fixing the instability threshold in terms of an effective Unruh temperature. Finally, we discuss how the stress-energy flux generated by Schwinger production may enter the generalized-entropy extremality condition, while leaving a complete backreaction and quantum-extremal-surface analysis for future work.
\end{abstract}

\maketitle

\section{Introduction}
Black holes are not merely solutions to Einstein’s equations, they are sharp probes into the quantum structure of spacetime. Hawking’s discovery that black holes radiate~\cite{Hawking:1975vcx,Unruh:1976db,Page:1976df} and Bekenstein’s realization that they carry entropy proportional to horizon area~\cite{Bekenstein:1973ur} placed them at the center of efforts to unify general relativity with quantum theory. Their thermodynamic properties~\cite{Hawking:1976de,Das:2001ic,Parikh:1999mf} have reshaped our understanding of entropy, information, and the microscopic underpinnings of geometry.

Lower-dimensional examples, most notably the charged BTZ black hole in $(2+1)$ dimensions, provide an analytically tractable arena for exploring these questions. In addition to the original BTZ construction and early boundary analyses~\cite{Banados:1992wn,Banados:1992gq,Carlip:1995qv}, the charged and spinning–charged sectors were worked out in detail by Cl\'ement and by Mart\'inez–Teitelboim–Zanelli~\cite{clement1996spinning,martinez2000charged}. More recent work has refined the definition of conserved charges in Einstein–Maxwell theory on AdS$_3$~\cite{perez2015conserved} and clarified the global structure and algebraic classification of the charged rotating family~\cite{maeda2024charged}. These spacetimes retain the essential thermodynamic and causal features of higher-dimensional black holes, yet are simple enough to allow exact computations in both the gravitational and matter sectors~\cite{Poojary:2022meo}.

In the near-extremal limit, the charged BTZ geometry develops a long AdS$_2\times S^1$ throat, a universal feature also seen in higher-dimensional near-extremal black holes~\cite{Cadoni:1999ja,Almheiri:2014cka}. This AdS$_2$ region is central to recent progress in AdS/CFT~\cite{Maldacena:1997re}, quantum chaos, nearly conformal quantum mechanics~\cite{Maldacena:2016hyu, Jensen:2016pah}, and entanglement dynamics in gravity~\cite{Almheiri:2019qdq,Penington:2019kki}. On the field theory side, it supports a clean definition of modular flow, thermal states, and vacuum correlations, key ingredients for holographic entanglement calculations.

Near-extremal throats also provide a well-controlled setting for semiclassical computations. Quantum corrections to the Bekenstein–Hawking entropy can be computed systematically using the quantum entropy function (QEF)~\cite{Sen2008,Sen2012,Sen:2012dw}, which incorporates one-loop effects from quantum fields in AdS$_2$. Techniques based on heat kernels and zeta functions~\cite{Fursaev1995,Frolov1997,Barrella2013,Banerjee2010} reveal universal logarithmic corrections and entanglement terms consistent with holographic expectations~\cite{Faulkner2014,Lewkowycz2013,Faulkner:2013ica}.

Here we focus on how the Schwinger effect, the spontaneous creation of charged pairs in an electric field~\cite{Schwinger:1951nm,Gibbons:1975kk,Chen:2012zn,Chen:2014yfa,Cai:2020trh,Chen:2020mqs}, manifests itself in the near-horizon region of a near-extremal charged BTZ black hole. Using a controlled expansion in the deviation from extremality, we construct an analytically solvable AdS$_2\times S^1$ background and solve the Klein–Gordon equation for a charged scalar field. The exact radial solution reveals the BF bound, the onset of instability, and the mean pair-production rate; in the appropriate limit, this reduces to the standard semiclassical Schwinger formula.\newline
We complement this with a tunneling treatment of Hawking radiation for charged particles, evaluating the imaginary part of the classical action in the WKB approximation to extract the Hawking temperature directly in throat coordinates. This allows us to compare the thermal and electric contributions to the emission spectrum and to interpret the instability threshold in terms of an effective Unruh temperature.\newline

\section{\label{sec:level2} Near-Horizon Geometry of the Near-Extremal Charged BTZ Black Hole}

We begin with the Einstein–Maxwell system in $(2+1)$ dimensions with a negative cosmological constant, which serves as the effective description of charged black holes in asymptotically AdS$_3$ spacetimes~\cite{Banados:1992wn,Carlip:1995qv,Banados:1992gq}. The action in which we will work is given by \cite{cadoni2008near}:
\begin{equation} \label{action1}
    S = \frac{1}{16 \pi G} \int d^3x \sqrt{-g} \left( R + \frac{2}{\ell^2} - 4G \pi F_{\mu\nu} F^{\mu\nu} \right),
\end{equation}
where $G$ is the three-dimensional Newton constant, $\Lambda=-1/\ell^2$ is the cosmological constant, with $\ell$ being the $\rm AdS$ length and $F_{\mu \nu}$ is the electromagnetic field strength tensor. In our case we consider the charged BTZ black hole with zero angular momentum using conventions in \cite{banados1993geometry}. A general solution for an electrically charged BTZ black hole given with the action \eqref{action1} is given by the three-dimensional line element \cite{Banados:1992gq, cadoni2008near} and with an electric field 
\begin{equation} \label{metric}
    ds^2 = -f(r)dt^2 + \frac{dr^2}{f(r)} + r^2 d\theta^2 \quad  \text{and} \quad F_{tr}= \frac{Q}{r}
\end{equation}
with lapse function:
\begin{equation} \label{lapse}
    f(r) = -8GM + \frac{r^2}{\ell^2} - 8 \pi GQ^2 \ln\left( \frac{r}{\ell} \right),
\end{equation}
where $M$ is ADM mass. The extremal limit corresponds to the case where inner and outer horizon coincide \cite{cadoni2008near} $r_+=r_-=\gamma$, i.e. when the lapse function \eqref{lapse} and its derivative vanish at the extremal radius.\newline
By setting $f^\prime(\gamma)=0$ we find that:
\begin{equation}
\gamma = 2 |Q| \ell \sqrt{\pi G}.
\end{equation}
In the remainder we take $Q>0$ for definiteness, so that $\gamma=2Q\ell\sqrt{\pi G}$ and the near-horizon electric field in Eq.~\eqref{fexpansion} is positive.
If we now substitute $\gamma$ into $f(\gamma)=0$ we find the extremal mass to be:
\begin{equation}
    M(\gamma) = \pi Q^2 \left( \frac{1}{2} - \ln{\frac{\gamma}{\ell}} \right) = \pi Q^2 \left( \frac{1}{2} - \ln{2 Q \sqrt{\pi G}} \right).
\end{equation}
What we are interested in this paper is the near horizon near extremal limit of \eqref{metric}. By performing a translation of the radial coordinate $r$ as 
\begin{equation} \label{expansion}
    r= \gamma +x,
\end{equation}
expanding the lapse function \eqref{lapse} and the gauge field in \eqref{metric} to second order in $x$ for $f$ and to first order for $F_{tx}$, we find that they take the following form \cite{cadoni2008near} :
\begin{equation}
   f(x)= \frac{2}{\ell^2}x^2 -8G\Delta M + \mathcal{O}(x^3) \quad F_{tx}= \frac{1}{ 2 \ell\sqrt{ \pi G }} + \mathcal{O}(x).
   \label{fexpansion}
\end{equation}
The full form of the near extremal near horizon metric is then given by \cite{cadoni2008near,myung2010entropy}:
\begin{equation} \label{metric2}
    ds^2 = -\left( \frac{2}{\ell^2}x^2 -8 G \Delta M \right) dt^2 +\frac{dx^2} {\frac{2}{\ell^2}x^2 -8 G \Delta M} + \gamma^2 d\theta^2,
\end{equation}
The two-dimensional factor in Eq.~\eqref{metric2} is the black-hole (Rindler) patch of AdS$_2$.  For $\Delta M>0$, the point $x=0$ lies between the inner and outer horizons and must not be identified with the event horizon.  The outer horizon is determined by

\begin{equation} \label{x+}
  \frac{2}{\ell^{2}}x_{+}^{2} - 8G\Delta M = 0 
  \quad \Rightarrow \quad
  x_{+} = \ell \sqrt{4G\Delta M},
\end{equation}
where $dx/dt \to 0$. 
The truncation in Eq.~\eqref{fexpansion} is controlled for $|x|\ll\gamma$; the AdS$_2$ ``boundary'' $x\to\infty$ used below should therefore be understood as the overlap region $x_+\ll x\ll\gamma$, which exists for $\Delta M\to0$ at fixed $Q$.
In the extremal limit $\Delta M \to 0$, the horizon radius collapses to $x_{+}=0$, and the lightcones pinch at the throat.\newline
The near-horizon near-extremal metric \eqref{metric2} therefore exhibits three essential features:
\begin{enumerate}
    \item \textbf{AdS$_2$ structure:} For $\Delta M = 0$, the $(t,x)$ sector reduces to
    \begin{equation}
      ds^2_{(2)} = -\frac{2}{\ell^2}x^2 dt^2 + \frac{\ell^2}{2x^2} dx^2 ,
    \end{equation}
    which is AdS$_2$ with curvature radius $R_A\equiv\ell/\sqrt{2}$; this symbol is used for the AdS$_2$ radius throughout.
    \item \textbf{Finite throat radius:} The transverse circle remains stabilized at radius
    \begin{equation}
      R_\theta = \gamma = 2\sqrt{\pi G}\,Q \,\ell ,
    \end{equation}
    independent of $x$ or $\Delta M$.
    \item \textbf{Near-extremal behavior:} For $\Delta M>0$, the static exterior is $x\geq x_+$, and excitations above extremality are encoded in $\Delta M$.
\end{enumerate}
In the near-horizon, near-extremal limit the gauge field reduces to a constant electric field. 
From the expansion \eqref{fexpansion} we read off
\begin{equation}
   F_{tx} = \frac{1}{2\sqrt{\pi G}\,\ell} + \mathcal{O}(x).
\end{equation}
A convenient gauge potential consistent with this is
\begin{equation}
   A_t(x) \approx  - \frac{x}{2\sqrt{\pi G}\,\ell} + \text{const.},
\end{equation}
which is approximately linear in the radial coordinate $x$. 
Thus the near-horizon geometry supports a constant AdS$_2$ electric field 
\begin{equation}
   E \equiv F_{tx} = \frac{1}{2\sqrt{\pi G}\,\ell},
\end{equation}
providing the background necessary for Schwinger pair production in AdS$_2$~\cite{Kim:2008xv,Chen:2014yfa,Kim:2015kna}.
This near-horizon limit provides a controlled and analytically tractable background for studying quantum field theory in curved spacetime. In the sections that follow, we use this geometry to analyze the instability triggered by charged pair production in the constant AdS$_2$ electric field, and to explore the resulting boundary quantum state and its modular structure.

\section{\label{sec:level3}Schwinger Pair Production in Near-Extremal BTZ Black Holes}

Quantum fluctuations in curved spacetime can spontaneously produce particle-antiparticle pairs, unless suppressed by the geometry or background fields. In black hole spacetimes, the event horizon acts as a causal barrier that separates such pairs, enabling the black hole to radiate all species of particles. This mechanism becomes particularly interesting for charged black holes.\newline
Extremal charged black holes, despite having zero Hawking temperature, can still support quantum emission through the Schwinger effect, a nonthermal tunneling process driven by the background electric field~\cite{Hawking:1975vcx,Schwinger:1951nm}. In the near-extremal regime, the geometry near the horizon develops an emergent $AdS_2 \times S^1$ throat. This structure allows for a precise analytic treatment of pair creation using methods developed in~\cite{Chen:2012zn,Chen:2020mqs,Cai:2020trh}. Related exact treatments of Schwinger production in AdS$_2$ include Refs.~\cite{Pioline:2005pf,Cai:2014qba}.\newline
We consider a minimally coupled, charged scalar field $\Phi$ with mass $m$ and charge $q$ described by the action
\begin{equation}
S = \int d^3 x \sqrt{-g} \left( -\frac{1}{2} D_\mu \Phi^* D^\mu \Phi - \frac{1}{2} m^2 \Phi^* \Phi \right),
\end{equation}
where $D_\mu = \nabla_\mu - i q A_\mu$ is the gauge-covariant derivative. The field $\Phi$ satisfies the Klein–Gordon equation:
\begin{equation} \label{kleingordon}
\left( \nabla_\mu - i q A_\mu \right) \left( \nabla^\mu - i q A^\mu \right) \Phi - m^2 \Phi = 0.
\end{equation}
We solve this equation using separation of variables:
\begin{equation}
\Phi(t,x,\theta) = e^{-i\omega t} H_L(\theta) R(x),
\end{equation}
where $H_L(\theta)$ is the eigenfunction of the angular Laplacian on $S^1$, satisfying $\partial_\theta^2H_L=-L^2H_L$, with $L\in\mathbb Z$.\newline
The radial equation in the near-horizon, near-extremal metric (cf. Section~\ref{sec:level2}) becomes:
\begin{align}
    &\left( \frac{2}{\ell^2}x^2 -8G \Delta M \right) R^{\prime \prime}(x) +  \left( \frac{4}{\ell^2}x \right) R^\prime(x) \notag\\
    &+ \left[ \frac{(\omega -qEx)^2}{\frac{2}{\ell^2}x^2-8G \Delta M} -m^2 -\frac{L^2}{\gamma^2} \right] R(x)= 0 \label{radialeq}
\end{align}
We can write the radial equation \eqref{radialeq} as:
\begin{equation}
    R^{\prime \prime}(x) + P(x) R^\prime(x) + Q(x) R(x)= 0.
\end{equation}
Writing the radial function $R(x)$ as
\begin{equation}
    R(x)= \chi(x) \psi(x),
\end{equation}
where 
\begin{equation} \label{chifunction}
    \chi(x)= \exp{-\frac{1}{2} \int P(s) ds},
\end{equation}
the canonical radial equation is then given by 
\begin{equation} \label{psiradial}
    \psi^{\prime \prime} + V_{\rm eff} \psi = 0,
\end{equation}
where the effective potential is now given by:
\begin{equation}
    V_{\rm eff} =  Q(x)- \frac{1}{2} P^{\prime}(x) - \frac{1}{4}P^2(x).
\end{equation}
For compactness, define $\alpha=2/\ell^2$ and $\beta^2=8G\Delta M$, so that
$f(x)=\alpha x^2-\beta^2$. The functions $P(x)$ and $Q(x)$ and the derivative
$P'(x)$ are then given by:
\begin{align}
   & P(x) = \frac{f^\prime(x)}{f(x)}=  \frac{2 \alpha x}{\alpha x^2- \beta^2} = \frac{\frac{4x}{\ell^2}}{\frac{2x^2}{\ell^2}-8G\Delta M}, \\
    &Q(x)=  \frac{(\omega-qEx)^2}{\left(\frac{2x^2}{\ell^2} -8G \Delta M\right)^2} - \frac{m^2+ \frac{L^2}{\gamma^2}}{ \frac{2x^2}{\ell^2}- 8 G\Delta M}, \\
    & P^\prime(x)=  \frac{-2 \alpha \left(\alpha x^2 +\beta^2\right)}{\left( \alpha x^2-\beta^2 \right)^2}.
\end{align}
In particular, the terms generated by removing the first derivative combine as
\begin{equation}
 -\frac{1}{2}P'(x)-\frac{1}{4}P(x)^2
 =\frac{\alpha\beta^2}{(\alpha x^2-\beta^2)^2}.
\end{equation}
It is useful to write the metric function in terms of a single AdS$_2$ radius,
\begin{equation}
 R_A^2\equiv\frac{\ell^2}{2},\qquad
 f(x)=\frac{x^2-x_+^2}{R_A^2},\qquad
 x_+^2=4G\Delta M\,\ell^2 ,
\end{equation}
where $R_A$ will be used throughout for the AdS$_2$ radius.  The effective
mass after reduction on the circle is
\begin{equation}
 m_{\rm eff}^2=m^2+\frac{L^2}{\gamma^2}.
\end{equation}
The metric is precisely the black-hole (Rindler) patch of AdS$_2$ with a
constant electric field.  Consequently, the full radial equation is exactly
solvable; no large-$x$ expansion is required.  This point is important because
the outer horizon is at $x=x_+$, whereas $x=0$ lies between the two horizons.
Thus a solution obtained by truncating in $x_+/x$ cannot consistently be used
to impose a horizon condition.

Introduce
\begin{equation}
 y=\frac{x-x_+}{x+x_+},\qquad 0\leq y<1,
\end{equation}
so that $y=0$ is the outer horizon and $y\to1$ is the AdS$_2$ boundary, and
define
\begin{equation}
 \widetilde\kappa=\frac{\omega R_A^2}{x_+},\qquad
 \kappa=qER_A^2,\qquad
 b=\sqrt{\kappa^2-m_{\rm eff}^2R_A^2-\frac14}.
 \label{exact_parameters}
\end{equation}
The two independent exact solutions may be chosen as
\begin{align}
 R_{\rm in}={}&y^{-\frac{i}{2}(\widetilde\kappa-\kappa)}
 (1-y)^{\frac12+ib}\nonumber\\
 &\times {}_2F_1\!\left(\frac12+i(b+\kappa),
 \frac12+i(b-\widetilde\kappa);
 1-i(\widetilde\kappa-\kappa);y\right),\label{Rin}\\
 R_{\rm out}={}&y^{+\frac{i}{2}(\widetilde\kappa-\kappa)}
 (1-y)^{\frac12+ib}\nonumber\\
 &\times {}_2F_1\!\left(\frac12+i(b+\widetilde\kappa),
 \frac12+i(b-\kappa);
 1+i(\widetilde\kappa-\kappa);y\right).\label{Rout}
\end{align}
These are respectively ingoing and outgoing at the future outer horizon,
as follows directly from the tortoise behavior
$y^{\mp i(\omega-qEx_+)R_A^2/(2x_+)}$.
At the AdS$_2$ boundary the two falloffs are
\begin{equation}
 R(x)\sim C_+x^{-1/2-ib}+C_-x^{-1/2+ib}.
\end{equation}
Analytically continuing $R_{\rm in}$ from $y=0$ to $y=1$ gives
$C_+\propto\Gamma(-2ib)/[\Gamma(\frac12-i(b+\widetilde\kappa))\Gamma(\frac12-i(b-\kappa))]$
and $C_-\propto\Gamma(2ib)/[\Gamma(\frac12+i(b+\kappa))\Gamma(\frac12+i(b-\widetilde\kappa))]$. With the flux convention used below, $x^{-1/2-ib}$ carries flux toward decreasing $x$ and is therefore the incident boundary mode, whereas $x^{-1/2+ib}$ is the reflected/outgoing boundary mode. The Bogoliubov connection coefficients then give the mean number of produced pairs \cite{Chen:2012zn,Kim:2015kna,Cai:2020trh}
\begin{equation}
 {\cal N}=
 \frac{\sinh(2\pi b)\,
       \sinh[\pi(\widetilde\kappa-\kappa)]}
      {\cosh[\pi(\kappa+b)]\,
       \cosh[\pi(\widetilde\kappa-b)]}.
 \label{meannumberofpairs}
\end{equation}
This expression follows from the conserved Klein--Gordon flux
\begin{equation}
 D=i\sqrt{-g}\,g^{xx}
 \left(\Phi D_x\Phi^*-\Phi^*D_x\Phi\right)
\end{equation}
and is the exact connection formula within the decoupled near-extremal throat.
For $\widetilde\kappa>\kappa$, flux conservation between the boundary and the future horizon is
$|D_{C_+}|=|D_{C_-}|+|D_{\rm hor}|$. Hence $C_+$ is the incident mode and $C_-$ is the reflected/outgoing mode, with the reflected flux smaller than the incident flux in this emission sector. The quantity ${\cal N}=|\beta|^2$ is the Bogoliubov mean number, while ${\cal N}/(1+{\cal N})$ is the corresponding transmission probability.
Moreover, since $T_H=x_+/(2\pi R_A^2)$ [see Eq.~\eqref{Hawking_temperature_S}],
\begin{equation}
 2\pi(\widetilde\kappa-\kappa)=\frac{\omega-qEx_+}{T_H},
\end{equation}
so the factor $\sinh[\pi(\widetilde\kappa-\kappa)]$ carries precisely the
Boltzmann factor of the gauge-covariant horizon energy
$\omega+qA_t(x_+)$ that appears in the tunneling analysis (Section IV). This provides an independent consistency check between the two
calculations.
Equation~\eqref{meannumberofpairs} is written for the emission sector
$\widetilde\kappa>\kappa$, for which the flux ratio is positive. Modes in the
complementary superradiant sector require an interchange of incoming and
outgoing fluxes and cannot be obtained by extending this expression without
changing the flux prescription.
In the extremal/low-temperature limit
$\widetilde\kappa\to\infty$, it reduces to
\begin{equation}
 {\cal N}_{\rm ext}=
 \frac{\sinh(2\pi b)}{\cosh[\pi(\kappa+b)]}
 e^{\pi(b-\kappa)} .
\end{equation}
This reproduces the extremal AdS$_2$ result of Refs.~\cite{Pioline:2005pf,Chen:2012zn,Cai:2014qba}, but it is now obtained as a controlled
limit of the exact hypergeometric solution rather than by imposing a boundary
condition outside the domain of a large-$x$ expansion.

\begin{figure}[ht]
  \centering
  \includegraphics[width=0.55\textwidth]{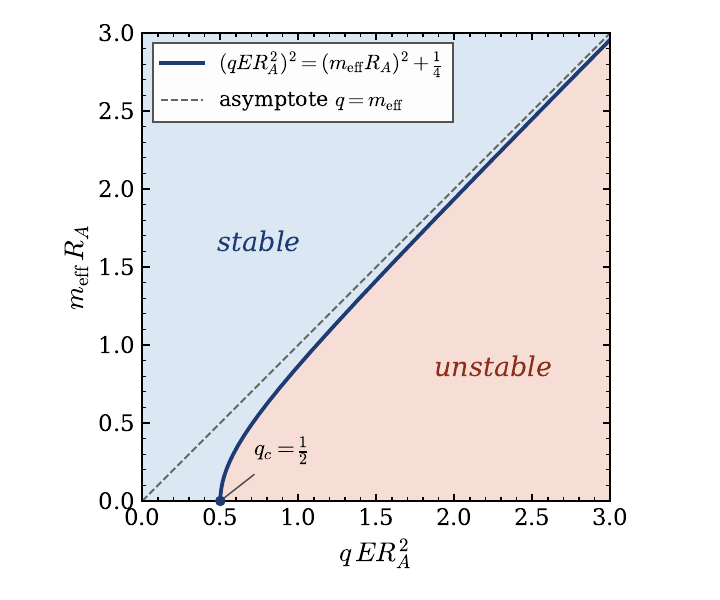}
  \caption{Stability diagram for a charged scalar in the AdS$_2$
  throat. The solid curve is the charged BF bound, $(qER_A^2)^2 = (m_{\rm eff}R_A)^2 + \tfrac14$. Above it (blue) the effective
  AdS$_2$ mass satisfies the BF bound and the mode is stable; below it (red)
  the bound is violated and the scalar is unstable. The dashed line is the
  large-mass asymptote $qER_A^2 = m_{\rm eff}R_A$. Even a massless field
  becomes unstable for $qER_A^2 > 1/2$.}
  \label{Stabilityregion}
\end{figure}

The charged AdS$_2$ BF combination \cite{Breitenlohner:1982bm,Breitenlohner:1982jf,Faulkner:2009wj} is most transparently defined as
\begin{equation}
 \Xi\equiv m_{\rm eff}^2R_A^2-(qER_A^2)^2,
 \qquad
 b=\sqrt{-\Xi-\frac14}.
 \label{BFbound}
\end{equation}
Stability requires
\begin{equation}
 (qER_A^2)^2\leq(m_{\rm eff}R_A)^2+\frac14,
\end{equation}
whereas Schwinger production is possible when this inequality is violated and
$b$ is real.  Equivalently,
\begin{equation}
 |q|_{\rm crit}=\frac{\sqrt{(m_{\rm eff}R_A)^2+1/4}}{|E|R_A^2},
\end{equation}
and 
\begin{equation}
 |E|_{\rm crit}=\frac{\sqrt{(m_{\rm eff}R_A)^2+1/4}}{|q|R_A^2}.
\end{equation}
The boundary dimensions (with standard and alternate quantization as in Ref.~\cite{Klebanov:1999tb}) are
\begin{equation}
 \Delta_\pm=\frac12\pm
 \sqrt{\frac14+m_{\rm eff}^2R_A^2-(qER_A^2)^2}.
 \label{scalingdim.}
\end{equation}
On the stable side (above the BF bound in mass), $\Delta_\pm$ are real.
The complementary-series terminology applies only when
$0<\Delta_\pm<1$; for $\Delta>1$ the representation is in the discrete
series.  At threshold the roots merge at $\Delta_\pm=1/2$, while on the
unstable side $\Delta_\pm=1/2\pm ib$ and the boundary modes are oscillatory.
Figure~\ref{Stabilityregion} uses $E=R_A=1$.

An interesting simplification occurs in the $s$-wave sector, $L=0$, when the
subleading $1/4$ contribution in the AdS$_2$ instability condition is neglected.
In this approximation, the condition for Schwinger pair production,
\begin{equation}
(q E R_A^2)^2>m^2R_A^2,
\end{equation}
reduces to
\begin{equation}
\frac{q}{m}>\frac{1}{E R_A}.
\end{equation}
For the near-horizon charged-BTZ geometry,
and therefore
\begin{equation}
E R_A=\frac{1}{2\sqrt{2\pi G}}.
\end{equation}
Hence the critical charge-to-mass ratio is
\begin{equation}
\left(\frac{q}{m}\right)_{\rm crit}
=2\sqrt{2\pi G},
\end{equation}
and the Schwinger instability requires
\begin{equation}
\frac{q}{m}>2\sqrt{2\pi G}.
\end{equation}
Remarkably, the AdS scale $\ell$ and the black-hole charge $Q$ cancel from the
final expression. Thus, within the $L=0$ near-horizon approximation, the
critical charge-to-mass ratio is a constant determined only by Newton's
constant. This universal form is reminiscent of weak-gravity-type
charge-to-mass bounds, although here it follows directly from the AdS$_2$
Schwinger instability condition rather than from the Weak Gravity Conjecture. The last equation indicates a Schwinger instability: probe species can satisfy the local near-horizon pair-production criterion. However, quantum-gravity models of charged black holes, such as T-duality-inspired black holes, may avoid this instability \cite{Chen:2026nna}.

\subsection{Semiclassical Schwinger exponent}

For $b,\kappa\gg1$, the extremal factor has the WKB form
\begin{equation}
 {\cal N}_{\rm ext}\simeq e^{-2\pi(\kappa-b)}.
\end{equation}
Expanding at large $R_A$ while holding the physical electric field $E$ fixed,
\begin{equation}
 2\pi(\kappa-b)
 =\frac{\pi m_{\rm eff}^2}{qE}
 +\frac{\pi}{4qER_A^2}\left(1+\frac{m_{\rm eff}^4}{q^2E^2}\right)+O(R_A^{-4}),
\end{equation}
and therefore
\begin{equation}
 {\cal N}_{\rm ext}\longrightarrow
 \exp\!\left[-\frac{\pi m_{\rm eff}^2}{qE}\right].
\end{equation}
This flat-space limit is formal within the charged-BTZ family unless the
parameters are scaled so that $E$ remains finite.  Since
$E=1/(2\sqrt{\pi G}\,\ell)$ and $R_A=\ell/\sqrt2$, this requires
$G\propto\ell^{-2}$.  We state this scaling explicitly and do not regard the
limit as one taken at fixed three-dimensional Newton constant.  Notice also
that the exact finite-temperature correction is encoded by
$\widetilde\kappa=\omega/(2\pi T_H)$ through Eq.~\eqref{meannumberofpairs};
no spurious $T_H^2/\omega^2$ term from a nonuniform large-$x$ expansion is
needed.

\section{Quantum tunneling of charged scalar particles}\label{sec:tunneling}
Let us elaborate and consider in more detail the Hawking radiation in terms of the tunneling  approach for the case of charged scalar particle in the spacetime \eqref{metric2}. The Klein-Gordon equation reads
\begin{equation}
\frac{1}{\sqrt{-g}}(\partial_{\mu}-\frac{i q}{\hbar}A_{\mu})\left( \sqrt{-g}g^{\mu\nu}(\partial_{\mu}-\frac{i q}{\hbar}A_{\mu})\Phi\right) -\frac{m^{2}}{\hbar^{2}}\Phi=0 \ ,  
\end{equation}
where $m$ and $q$ are the mass and charge of the particle, respectively. We will make use of the WKB ansatz,
\begin{equation}
\Phi=\exp\left(\frac{i}{\hbar}S(t,x,\theta)\right) \ ,
\end{equation}
In general, one can expand the action $S(t,x,\theta)$ in powers of $\hbar$ as follows
\begin{equation}
    S(t,x,\theta)=S_{0}(t,x,\theta)+\sum_{i=1} \hbar^i S_{i}(t,x,\theta).
\end{equation}
Taking into the consideration the symmetries of the metric by corresponding Killing vectors $(\partial/\partial t)^{\mu}$ and
$(\partial/\partial\theta)^{\mu}$, the action takes the form
\begin{equation}
S_0(t,x,\theta)=-\mathbb{E}t+W(x)+L\theta,
\end{equation}
where $\mathbb{E}$ is the energy of the particle and $L$ denotes the angular momentum of the particle corresponding to the angle $\theta$. 
As noted in \cite{Parikh:1999mf}, the WKB approximation can be justified due to the following argument: It is expected that the typical wavelength of the radiation to be of the order of the size of the horizon, however, when the outgoing wave is traced back towards the horizon, its wavelength as measured by static observers is increasingly blue-shifted. Near the horizon, the radial wavenumber approaches infinity and the point particle, or WKB, approximation is justified. From the above equations in the leading order terms in $\hbar$,  we obtain the following equation 
\begin{eqnarray}\notag
  &-&\left(\frac{2x^2}{\ell^2}-8 G \Delta M \right) \left(\frac{d }{d x} W(x)\right)^2 +\frac{(\mathbb{E}+q A_t)^2}{\left(\frac{2x^2}{\ell^2}-8 G \Delta M \right)}\\
  &-&\left(m^2+\frac{L^2}{\gamma^2}\right)=0
\end{eqnarray}
from where we get the radial part $W(x)$ as follows
\begin{equation}
    W_{\pm}=\pm \int  \frac{\sqrt{(\mathbb{E}+q A_t)^2- (\frac{2x^2}{\ell^2}-8 G \Delta M)\left(m^2+\frac{L^2}{\gamma^2}\right)}}{\frac{2x^2}{\ell^2}-8 G \Delta M}\,\,dx
\end{equation}
To evaluate this integral, we first define the function $\mathcal{F}(x)=\frac{2x^2}{\ell^2}-8 G \Delta M $, and we consider a series expansions around the outer horizon $x_+$, namely
\begin{eqnarray}
    \mathcal{F}(x)\simeq \mathcal{F}'(x)|_{x=x_+} (x-x_+)
\end{eqnarray}
where $\mathcal{F}'(x)|_{x=x_+} = 4 x_+ /\ell^2$; higher-order terms in $(x-x_+)$ do not contribute to the residue at the pole. In order to find the Hawking temperature, we now make use of the equation
\begin{equation}
\lim_{\epsilon \to 0} \text{Im}\frac{1}{x-x_+\pm i \epsilon }=\mp\pi\,\delta(x-x_+)~,
\end{equation}
in this way we find 
\begin{equation}
\text{Im}W_{\pm}=\pm \frac{ \tilde{\mathbb{E}}_H\, \pi }{\mathcal{F}'(x)|_{x=x_+}}~,
\end{equation}
where $\tilde{\mathbb{E}}_H=\mathbb{E}+q A_t(x_+)$ is the gauge-covariant horizon energy, with the additive constant in $A_t$ fixed by the chosen asymptotic gauge. Using $p_x^{\pm}=\pm \partial_x W_{\pm}$, for the total tunneling rate we get
\begin{align}\notag
\Gamma &=\exp\left(\frac{1}{\hbar}\text{Im} (\tilde{\mathbb{E}}_H \Delta t^{\rm out,in})-\frac{1}{\hbar}\text{Im} \oint p_{x} \mathrm{d}x\right)\\
&=\exp\left(-\frac{4 \tilde{\mathbb{E}}_H \pi }{\mathcal{F}'(x)|_{x=x_+}}\right) \ .
\end{align}
In the last equation we have also added a temporal part contribution due to the connection of the interior region and the exterior region of the spacetime metric via $t \to t - i \pi /\mathcal{F}'(x_+)$. Finally, 
one can obtain the Hawking temperature once we use the Boltzmann factor $\Gamma=\exp(-\tilde{\mathbb{E}}_H/T_H)$, and set $\hbar$ to unity. That results in
\begin{equation}\label{Hawking_temperature_S}
    T_H=\frac{ \mathcal{F}'(x)|_{x=x_+}}{4 \pi }=\frac{x_+}{\pi \ell^2}=\frac{ x_+}{2\pi R_A^2}.
\end{equation}
In the case of an extremal black hole we have $x_+ = \ell \sqrt{ 4 G  \Delta M}  \to 0$, since $\Delta M \to 0$, which means that $T_H=0$. Formally, setting $x_+=R_A$ in the throat metric gives $T_H=1/(2\pi R_A)$; however, this choice need not lie within the controlled near-extremal regime $x_+\ll\gamma$ and is not used in the subsequent analysis.

\subsection{A thermal-like nature of Schwinger effect}
The spontaneous discharge of a charged black hole results from both Hawking radiation and the Schwinger effect. In this direction, it is interesting to mention a possible link and analogy between Hawking radiation and the Schwinger effect. Let us use the Schwinger tunneling rate and rewrite it as follows
\begin{eqnarray}
    \Gamma_S \sim \exp\left(-\frac{\pi m^2}{q \mathcal{E}}  +\rm corrections\right),
\end{eqnarray}
Let us assume that $\mathcal{E}=\mathcal{E}_{\rm local}$ is some local measured electric field, then we further assume a Boltzmann factor $\Gamma=\exp(-m/T_S)$, i.e., we may write a thermal-like temperature, which is found to be
\begin{eqnarray}
   T_S\sim\frac{q \mathcal{E}}{ \pi m},
\end{eqnarray}
in units $c=\hbar=k_B=1$. Since $ q \mathcal{E}=m a$, where $a$ is some proper accelerations, we get
\begin{eqnarray}
    T_S\sim \frac{a}{\pi} \sim 2\, T_U,
\end{eqnarray}
where we have used the Unruh-like temperature given by $T_U=a/(2\pi)$. This is an interesting case as it shows that the thermal-like temperature  $T_S$ of the Schwinger effect which can be viewed as a Unruh-like temperature. Note that this is just an analogy and the correspondence is not exact. On the other hand, we can write the Hawking temperature given by Eq. \eqref{Hawking_temperature_S} in terms of surface gravity 
\begin{eqnarray}
     T_H=\frac{\kappa_{\rm s}}{2 \pi},
\end{eqnarray}
where $\kappa_{\rm s}=x_+/R_A^2$ is the surface gravity (not to be confused with $\kappa=qER_A^2$ of Sec.~\ref{sec:level3}). For the metric used here, the electric field measured in an orthonormal frame is
\begin{equation}
 E_{\hat t\hat x}=e_{\hat t}^{\ t}e_{\hat x}^{\ x}F_{tx}
 =\frac{1}{\sqrt{f}}\sqrt{f}\,F_{tx}=F_{tx}=E,
\end{equation}
so no additional factor $\sqrt{-g_{tt}}$ multiplies the local electric field. Introduce
\begin{equation}
 \bar m=\sqrt{m_{\rm eff}^2+\frac{1}{4R_A^2}},
 \qquad \bar T_U=\frac{qE}{2\pi\bar m}.
\end{equation}
The exact BF threshold is then $\bar T_U>1/(2\pi R_A)$. If instead
$T_U=qE/(2\pi m_{\rm eff})$, the equivalent condition is
\begin{equation}
 T_U^2>\frac{1}{4\pi^2R_A^2}
 +\frac{1}{16\pi^2m_{\rm eff}^2R_A^4}.
\end{equation}
Thus $T_U\gtrsim1/(2\pi R_A)$ is only the semiclassical approximation
$m_{\rm eff}R_A\gg1$. In this direction, it is useful to introduce the thermal-like Schwinger temperatures \cite{Kim:2008xv,Cai:2020trh}
\begin{align} \label{temperatures}
T_S &= \bar T_U + \sqrt{\bar T_U^2 - \frac{1}{4 \pi^2 R_A^2}}, \\
\bar{T}_S &= \bar T_U - \sqrt{\bar T_U^2 - \frac{1}{4 \pi^2 R_A^2}},
\end{align}
The above equations of temperature \eqref{temperatures} should be understood as effective temperatures characterizing the exponential suppression of tunneling processes, rather than as genuine equilibrium temperatures associated with a thermal density matrix. The curvature-shifted effective temperature $\bar T_U$ entering these expressions characterizes the acceleration of charged excitations in the near-horizon electric field, while the $\rm AdS_2$ curvature scale $R_A$ sets the geometric threshold. The square-root structure encodes the competition between electric-field-induced acceleration, which enhances pair production, and spacetime curvature, which suppresses it. When the argument of the square root becomes negative, the effective temperature becomes complex, signaling the absence of a tunneling instability. Thus, Eqs. \eqref{temperatures} provide a useful semiclassical parametrization of the Schwinger rate in thermal-like form, without implying true thermality or detailed balance.
In fact, since $\bar m/T_S=2\pi(\kappa-b)$ and $\bar m/\bar T_S=2\pi(\kappa+b)$, the exact extremal result of Sec.~\ref{sec:level3} can be written identically as
\begin{equation}
 {\cal N}_{\rm ext}=\frac{e^{-\bar m/T_S}-e^{-\bar m/\bar T_S}}{1+e^{-\bar m/\bar T_S}},
\end{equation}
which is the AdS$_2$ form of Ref.~\cite{Cai:2014qba}; in the flat limit $R_A\to\infty$ one has $T_S\to 2\bar T_U$ and $\bar m\to m_{\rm eff}$, recovering $e^{-\pi m_{\rm eff}^2/(qE)}$.\newline
This thermal-like form shows a link between near-horizon acceleration (Unruh temperature) and tunneling suppression. The electric field plays a dual role: it lowers the tunneling barrier, facilitating pair production, but also sets the charged BF threshold in Eq.~\eqref{BFbound}, beyond which pair production becomes possible~\cite{Kim:2015kna,Chen:2019rtd,Cai:2020trh}. 

\section{Discussion and Outlook} \label{discussion}

We analyzed the near-horizon region of a near-extremal \emph{charged} BTZ black hole in a controlled AdS$_2\times S^1$ throat. In this background we solved the charged Klein–Gordon equation exactly in hypergeometric form within the decoupled AdS$_2$ black-hole patch, identified the BF-bound threshold, and computed the Schwinger pair-production rate, recovering the standard semiclassical limit. A WKB tunneling analysis done directly in throat coordinates isolates the charged Hawking channel and cleanly separates thermal from electric contributions, fixing the instability threshold via an effective Unruh temperature. \newline
The stress-energy flux of the produced pairs provides a natural source for a future semiclassical backreaction analysis. A complete derivation of the generalized-entropy variation and the location of a quantum extremal surface, however, requires the dilaton variation and the quantum matter entropy beyond the frozen-circle throat approximation. Our near-horizon metric agrees with prior presentations up to standard coordinate/gauge choices but is organized for the Schwinger analysis and near-extremal expansion.\newline
Our near-horizon AdS$_2\times S^1$ analysis is closely parallel to the Coulomb-impurity problem on AdS$_2\times S^1$ studied in~\cite{Aharony:2022ntz,Aharony:2023amq}. In that setup, a similar radial equation, BF-bound threshold, and defect-RG structure appear, and a screening cloud can extend macroscopically. We expect the same endpoint in the black-hole throat, up to (i) ingoing horizon boundary conditions/redshift and (ii) charge quantization and global constraints on the $S^1$. We hope to report elsewhere on the explicit parameter matching and its implications for charge quantization.

\subsection*{Acknowledgement}{We thank Z.Komargodski for pointing out the close analogy with the Coulomb impurity problem in AdS$_2 \times S^1$and for highlighting  Refs.~\cite{Aharony:2022ntz,Aharony:2023amq}.}

\appendix
\section{Entanglement and near-extremal thermodynamics}
We first clarify the limitation of applying the RT prescription directly
to the frozen-circle throat. On a constant-$t$ slice the induced metric is
\begin{equation} \label{slicedmetric}
ds^2_{\rm slice}=\ f^{-1}(x) dx^2 +\gamma^2\,d\theta^2.
\end{equation}
For curves $x(\theta)$, the length functional is
\begin{equation} \label{RTcurve}
\mathcal{L}[x(\theta)]
= \int d\theta\,\mathcal{I},\qquad \mathcal{I}\equiv\sqrt{\,\frac{1}{f(x)}\left(\frac{dx}{d\theta}\right)^2 + \gamma^2\,}.
\end{equation}
Because the integrand has no explicit $\theta$-dependence, the Beltrami identity gives:
\begin{equation}
\mathcal{B}=\mathcal{I}-x'\,\frac{\partial\mathcal{I}}{\partial x^\prime}
=\frac{\gamma^2}{\sqrt{\frac{x'^2}{f(x)}+\gamma^2}}
= {\rm const}.
\end{equation}
At any turning point $x^\prime(\theta_*)=0$ one has $\mathcal{I}(\theta_*)=\gamma$, hence the constant is $\gamma$ and the above equation forces
\begin{equation}
\frac{x^{\prime2}}{f(x)}=0 \quad\Rightarrow\quad x^\prime(\theta)=0.
\end{equation}
Thus any constant-$t$ extremal in this product geometry that possesses a radial turning point is forced to remain at constant radius:
\begin{equation}
x(\theta)=x_0, \qquad \mathcal{L}=\gamma\,\Delta\theta.
\end{equation}
Monotonic geodesics without a turning point are not excluded by this argument; in the proper radial coordinate $du=dx/\sqrt{f}$ they satisfy the flat-cylinder geodesic equation $u=a\theta+b$.
However, a constant-$x$ curve can connect endpoints anchored at the
cutoff $x=x_c$ only if it remains at $x=x_c$. It does not define a
nontrivial boundary-anchored RT surface that turns at an arbitrary bulk
radius. This degeneracy is a consequence of freezing
$g_{\theta\theta}=r^2$ to $\gamma^2$ in the strict throat limit. A genuine
boundary-interval RT calculation must retain
$g_{\theta\theta}=(\gamma+x)^2$ or use the full charged-BTZ geometry.
Nevertheless, the circumference of the extremal horizon correctly gives
\begin{equation}
S_{\rm BH}=\frac{2\pi\gamma}{4G}=\frac{\pi\gamma}{2G}.
\end{equation}
From our metric \eqref{metric2}, the outer horizon in the original BTZ coordinate is $r_+=\gamma+x_+$.  Therefore
\begin{eqnarray}\notag
 S_{\rm near-ex}&=&\frac{A}{4G}
 =\frac{2\pi(\gamma+x_+)}{4G}\\
&=&\frac{\pi\gamma}{2G}+\pi\ell\sqrt{\frac{\Delta M}{G}}
 =\frac{\pi\gamma}{2G}+\frac{\pi^2\ell^2}{2G}T_H.
\end{eqnarray}
Here $x_+=\ell\sqrt{4G\Delta M}$ and $T_H=x_+/(\pi\ell^2)$ are relations of the quadratic throat metric.  The result is dimensionless because $[G]=L$ and $[\Delta M]=L^{-1}$.
For the complete charged-BTZ lapse these expressions are the leading near-extremal terms. Expanding one order further gives
\begin{equation}
f(\gamma+x)=-8G\Delta M+\frac{2x^2}{\ell^2}-\frac{2x^3}{3\gamma\ell^2}+O(x^4),
\end{equation}
so the exact full-geometry horizon and temperature acquire relative corrections of order $x_+/\gamma$.
Accordingly, we do not identify
$\gamma\min(\Delta\theta,2\pi-\Delta\theta)/(4G)$ with a boundary-interval
entanglement entropy in the strict product throat.
The total generalized entropy is written as
\begin{equation}
S_{\rm gen}=\frac{A}{4G}+S_{\rm vN},
\end{equation}
where $S_{\rm vN}$ is the quantum von Neumann contribution from fields in the
near-horizon region. For a null cut whose local modular flow is generated by $\zeta=2\pi U\partial_U$, a schematic first-order modular-energy variation sourced by the Schwinger-produced pairs has the form
\begin{equation}
    \delta K_\zeta \sim 2\pi\int dU\,d\theta\,\gamma\,U\,\delta T_{UU},
\end{equation}
with the integration range and orientation fixed by the chosen null region. The formal extremality condition is
\begin{equation}
    \delta S_{\rm gen}
    =\frac{\delta A}{4G}+\delta S_{\rm vN}=0.
\end{equation}
The first law of entanglement can relate $\delta S_{\rm vN}$ to a
modular-energy variation in an appropriate reference state. Nevertheless,
the local equation $R_{UU}=8\pi G\,\delta T_{UU}$ and the position of a QES
do not follow from the flux expression alone. Their derivation requires a
specified null deformation, the varying circle radius (the two-dimensional
dilaton), the quantum state and its entropy, and the semiclassical Einstein
equations. We therefore regard $\delta K_\zeta$ only as schematic input for such a
future backreaction calculation and do not claim to locate a QES here.\newline
Within the quadratic throat approximation, Eq.~\eqref{x+} gives $x_+^2=4G\Delta M\ell^2$ and $T_H=x_+/(\pi\ell^2)$, hence
\begin{equation}
    \Delta M = \frac{\pi^2 \ell^2}{4G} T_H^2+O(T_H^3).
\end{equation}
The $O(T_H^3)$ notation emphasizes that, for the complete charged-BTZ geometry, the cubic term in $f(\gamma+x)$ corrects both the horizon relation and the surface gravity beyond leading near-extremal order.
One can show that in the canonical/Hamiltonian formulation, the generator $H[\xi]$ of a symmetry $\xi^a$ picks up a boundary term at spatial infinity. For the logarithmically asymptotic charged-BTZ solution, the identification of that boundary charge with the parameter $M$ depends on the chosen electromagnetic boundary conditions and renormalization prescription. At fixed $Q$ and within a fixed prescription we may therefore identify $\Delta E=\Delta M$ for variations about the same extremal background.\newline
At leading near-extremal order this is consistent with $d(\Delta M)=T_H\,dS_{\rm near-ex}$. Finally, the energy above extremality has the same universal quadratic scaling discussed in \cite{Iliesiu:2020qvm,Kudler-Flam:2023qfl}:
\begin{equation}
    \Delta E= \frac{T_H^2}{M_{\rm gap}}+O(T_H^3),
\end{equation}
where $M_{\rm gap}=\frac{4G}{\pi^2\ell^2}$ at leading order.

\bibliographystyle{apsrev4-2}
\bibliography{biblo}

\end{document}